\documentclass[runningheads]{llncs}
\usepackage{graphicx}

\usepackage{color,soul,booktabs,eurosym}

\begin{document}
    \title{Dark Patterns in Online Shopping:\\Of Sneaky Tricks, Perceived Annoyance and Respective Brand Trust}
\titlerunning{Dark Patterns in Online Shopping}
%

%
%

\author{Christian Voigt \and
Stephan Schl\"{o}gl\orcidID{0000-0001-7469-4381} \and
Aleksander Groth
}
%
\authorrunning{Voigt et al.}
%
\institute{MCI -- The Entrepreneurial School\\ Innsbruck, Austria\\
\email{stephan.schloegl@mci.edu}\\
\url{https://www.mci.edu} 
}

%
\maketitle              
\begin{abstract}
Dark patterns utilize interface elements to trick users into performing unwanted actions. Online shopping websites often employ these manipulative mechanisms so as to increase their potential customer base, to boost their sales, or to optimize their advertising efforts. Although dark patterns are often successful, they clearly inhibit positive user experiences. Particularly, with respect to customers' perceived annoyance and trust put into a given brand, they may have negative effects. To investigate respective connections between the use of dark patterns, users' perceived level of annoyance and their expressed brand trust, we conducted an experiment-based survey. We implemented two versions of a fictitious online shop; i.e. one which used five different types of dark patterns and a similar one without such manipulative user interface elements. A total of $n=204$ participants were then forwarded to one of the two shops (approx. $2/3$ to the shop which used the dark patterns) and asked to buy a specific product. Subsequently, we measured participants' perceived annoyance level, their expressed brand trust and their affinity for technology. Results show a higher level of perceived annoyance with those who used the dark pattern version of the online shop. Also, we found a significant connection between perceived annoyance and participants' expressed brand trust. A connection between participants' affinity for technology and their ability to recognize and consequently counter dark patterns, however, is not supported by our data.
\keywords{Dark Patterns \and Online Shopping \and Perceived Annoyance \and Brand Trust Scale \and Affinity for Technology.}
\end{abstract}
%
%
\section{Introduction}
Design patterns have the ability to significantly affect the way people interact with user interfaces and consequently how such interactions are perceived. Hence, there is a connection between design elements and whether or not they help users reach desired outcomes~\cite{sommerer2008art}. To this end, online shops increasingly utilize techniques which help trigger subconscious or potentially unintended customer behavior to support e-commerce goals (e.g., increase customer reach, boost sales numbers, optimize advertising efforts, etc.). These so-called \emph{dark patterns} aim at tricking users into unintended actions, such as for example the unwilling subscription to a newsletter or the accidental clicking on an ad\footnote{Online: https://darkpatterns.org [accessed: November 2\textsuperscript{nd} 2020]}. 

A detailed analysis of 680 out of the 10.000 most viewed websites in the United Kingdom, showed that over 88\% of them use at least some sort of dark patterns~\cite{nouwens2020dark}. Given, that on average an Internet user (between 16 and 64 years) spends approx. 6 hours and 40 minutes online per day, the superfluous use of dark patterns may be considered a common, although somewhat unethical, business practice. A practice not only to be found on e-commerce websites but also on smartphone apps (Note: a random selection of 240 apps available in the Google Play Store showed that 95\% of them use dark patterns).
While the impact of dark patterns on brand trust (i.e., how and how much dark patterns potentially harm perceived trust in a brand) has not yet been fully investigated, a negative impact on app ratings as well as usage has been identified~\cite{di2020ui}.
In order to increase our understanding within this field, our study aims to explore the use of dark patterns in a simulated e-commerce setting and how these are connected to peoples' perceived level of annoyance as well as the respective trust they put into brands when shopping online. Respective analyses were guided by the following research question:

\begin{quote}
\textit{``What is the connection between dark patterns, perceived level of annoyance and brand trust when shopping online?''}
\end{quote}

\noindent
We will begin with a brief discussion describing the emergence of software patterns and the accompanying rise of dark patterns highlighted by Section~\ref{sec:darkpatterns}. In  Section~\ref{sec:methodology} we outline our research design and respective methodology. Section~\ref{sec:results} presents our results and Section~\ref{sec:discussion} discusses their relevance and limitations. Finally, Section~\ref{sec:conclusion} concludes and proposes future research directions.

\section{The Rise of Dark Patterns}\label{sec:darkpatterns}
Originally applied in structural engineering and architecture~\cite{alexander1977pattern}, design patterns were popularized within software development in the early 1990s~\cite{gamma1995design}. Ever since, their goal has been to improve overall software quality and to offer template solutions to reoccurring architectural problems. However, the proliferation of these `good design practices' was accompanied by a similar strong appearance of \emph{anti-patterns} (i.e, bad design practices), as well as \emph{dark patterns} which are described as \textit{``[...] instances where designers use their knowledge of human behavior (e.g. psychology) and the desires of end users to implement deceptive functionality that is not in the user’s best interest''}~\cite[p. 1]{gray2018dark}~\cite{bosch2016tales}. 
As for the latter, three fields of application are identified, which have seen a particularly high level of respective adoption~\cite{narayanan2020dark}.
First, in retail, where customers are being increasingly exposed to deceptive practices such as \emph{psychological pricing} or \emph{false advertising}. Second, in sales management, where through \emph{growth hacking} companies have been using a combination of big data analysis, marketing and deceiving design practices to increase product adoption. And third, in consumer research and public policy, where \emph{nudging} has become the de facto standard for affecting and consequently changing people's behaviour. From a user (or customer) perspective those practices may be deemed unethical, and with respect to the processing of personal data even unlawful.
Ever since the introduction of the General Data Protection Regulation (GDPR) by the European Union in April 2016, the processing of personal data has been legally restricted.
Outlining that the consent to process data regards: 
\begin{quote}
\textit{``[...] any freely given, specific, informed and unambiguous indication of the data subject’s wishes by which he or she, by a statement or by a clear affirmative action, signifies agreement to the processing of personal data relating to him or her.''}
~\cite[Article 4(8)]{eugdpr}
\end{quote}

\noindent
the GDPR regulates e.g. the implementation and utilization of so-called cookies on websites, pushing for particularized user consent. Yet, a recent study in the UK clearly shows that respective compliance is often missing: Out of 680 investigated websites only 11.8\% offered (1) \textbf{explicit consent} where (2) \textbf{accepting all} is \textbf{as easy as rejecting all}, and (3) selection \textbf{boxes} are \textbf{not pre-ticked}~\cite{mohan2019analyzing}. To this end, the concepts \emph{privacy dark patterns} and \emph{privacy dark strategies} are of particular interest, as they describe strategies which focus exclusively on the exploitation of personal data, thereby underlining the great relevance such unethical practices still have in e-commerce~\cite{bosch2016tales}.

Previous work has outlined various concepts on how to integrate ethical considerations into the design of and consequent interaction with modern user interfaces. These practices usually center around value approaches such as Value Sensitive Design (VSD)~\cite{friedman1996value} or Value Levers~\cite{shilton2013values}. All of them have in common that they aim for human values to be prioritized in the (technical) design processes~\cite{friedman2019value,shilton2013values}. While the use of dark patterns clearly violates such values, and thus may be considered unethical, one might argue that it is not only ethical concerns which are at play here. Although companies might increase their benefits by tricking customers into unwilling behavior, these practices often lead to a negative user experiences. Additionally, they may affect the trust customers put into a given brand, which eventually might harm an organization's reputation. E-commerce companies may thus be willing to refrain from using dark patterns if it can be shown that the negative consequences regarding long-term brand trust outweigh respective gains in short-term sales.

\section{Methodology}\label{sec:methodology}
In order to investigate a potential connection between brand trust and the use of dark patterns, we used an experiment based on a between-subject design in which two versions of a fictitious online shop served as stimuli. One shop (i.e., \textsc{dark}) used five different dark patterns based on the categorization by Gray and colleagues~\cite{gray2018dark}, the other (i.e., \textsc{clean}) was free of such manipulative interface elements (cf. Section~\ref{sec:stimulus}).

\subsection{Stimuli}\label{sec:stimulus}
\begin{figure}[ht]
    \centering
    \includegraphics[width=100mm]{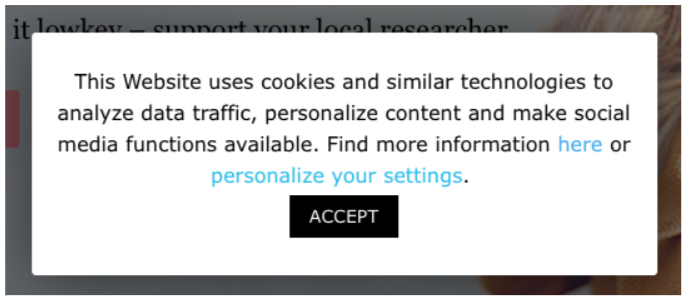}
    \caption{Cookie consent banner using the \emph{Forced Action} pattern}
    \label{fig:banner_dark}
\end{figure}

As outlined above, we used two different versions of the same online shop to investigate upon perception differences connected to the appearance of dark patterns. Following, we describe the patterns implemented by the \textsc{dark} version of our shop and compare them to the respective interface elements used by the \textsc{clean} version:\\[1\baselineskip] 
\textbf{\emph{Forced Action:}} 
Forced action is one of the first and most commonly dark patterns users will encounter when visiting a website. According to Gray et al., \textit{forced action} describes a situation where \textit{``users are required to perform a specific action to access (or continue to access) a specific functionality.''}~\cite{gray2018dark}. When encountering such a pattern, users are given little choice but to follow a pre-set path. An often implemented type of forced action is referred to as \emph{Privacy Zuckering} and is found in cookie consent banners. The goal is to make users share their information in ways they do not mean to. This is achieved by making mandatory privacy settings intentionally complicated and/or incomprehensible to use~\cite{bosch2016tales}\cite{gray2018dark}. For example, the \textsc{dark} version of our online shop uses a banner to outline cookie and privacy settings. It appeared in the center of the website and blocked all further access to the other navigational elements. In order to close it, users were visually drawn towards the already and only pre-selected ``Accept'' button (note: the accept button was the most prominent visual interface element due to its black background), thereby agreeing to activate all cookies and so accepting full user tracking (cf. Fig.~\ref{fig:banner_dark}). 

\begin{figure}[ht]
    \centering
    \fbox{\includegraphics[width=100mm]{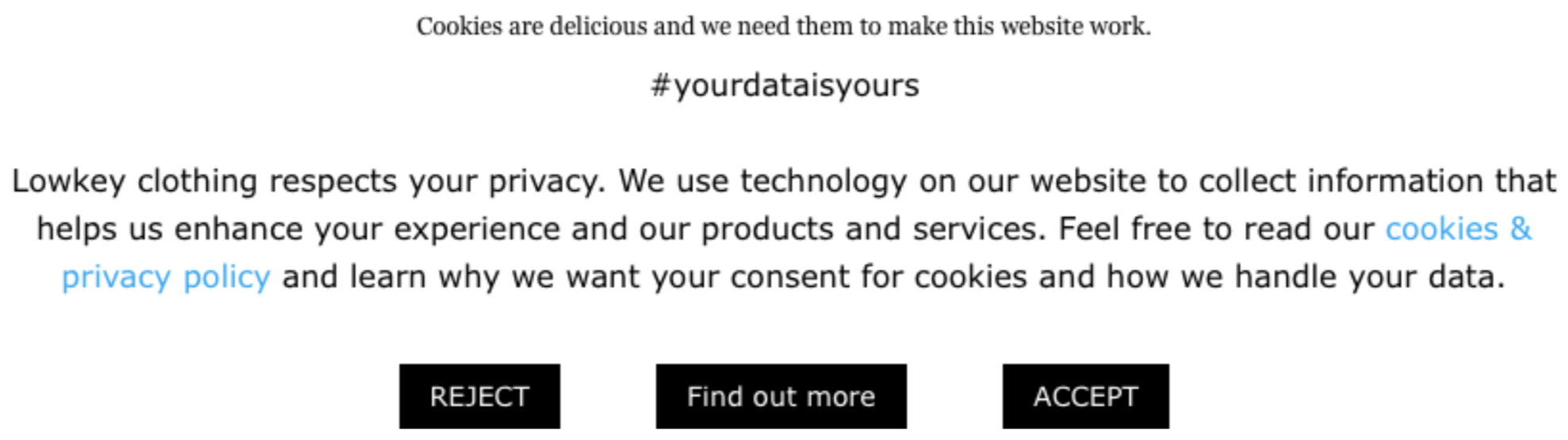}}
    \caption{GDPR compliant cookie consent banner}
    \label{fig:banner_clean}
\end{figure}

\noindent
To adjust these settings and reject all unnecessary tracking, users had to select the ``personalize your settings'' option presented as a text-link (note: we used light blue colored text with intentionally low contrast so that this option would be significantly less obvious than the accept button). It should be noted that this type of cookie banner is not compliant with GDPR rules, yet still found on many websites. The \textsc{clean} version of our online shop showed a cookie banner which did not push users towards accepting unwanted settings, and thus may be considered GDPR compliant (cf. Fig.~\ref{fig:banner_clean}).

\begin{figure}[ht]
    \centering
    \includegraphics[width=100mm]{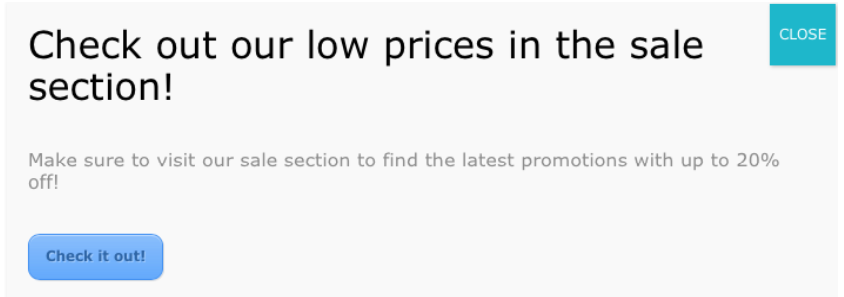}
    \caption{Advertisement Pop-up representing \emph{Nagging}}
    \label{fig:Nagging}
\end{figure}

\noindent
\textbf{\emph{Nagging:}} 
Nagging uses scheduled intrusions such as pop-ups to repeatably disrupt normal interaction workflows~\cite{chromik2019dark}. Users may develop militant feeling against these nagging elements, which may go as far as to trigger a purchase for the sole purpose of skipping forward~\cite{gray2018dark}\cite{lewis2014irresistible}. Following this approach, the \textsc{dark} version of our online shop presented pop-ups on every single product page, inviting users to visit the sales section. One had to either close the pop up or visit the sales section in order to proceed. All other navigation elements were blocked (cf. Fig.~\ref{fig:Nagging}). Additionally, a similar pop-up was triggered when users reviewed the shopping cart, inviting them once more to visit the shop's sales section. On the other hand, the \textsc{clean} version of our online shop was completely free of such pop-ups. 

\begin{figure}[ht]
    \centering
    \includegraphics[width=40mm]{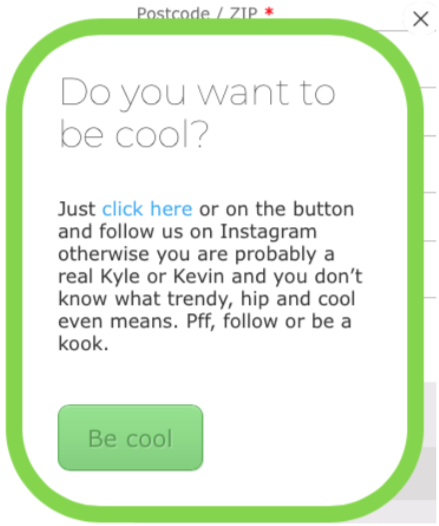}
    \caption{Pop-up representing \emph{Confirmshaming}}
    \label{fig:confirmshaming}
\end{figure}

\noindent
\textbf{\emph{Confirmshaming:}} 
When users of the \textsc{dark} version of our online shop eventually proceeded to the checkout, they were asked to enter their name and shipping address. Here we implemented the so-called confirmshaming pattern, which in our case was represented by a pop-up window trying to convince users to ``Be cool'' (cf. Fig.~\ref{fig:confirmshaming}). 

\noindent
The respective button opened a new browser tab, showing our school's Instagram page. According to Gray~\cite{gray2018dark}, the goal of the confirmshaming pattern is to use dedicated interface elements such as specific language, sound, color or style to convey emotions and consequently incite users to perform an originally unintended action.\\[1\baselineskip]

\begin{figure}[ht]
    \centering
    \includegraphics[width=75mm]{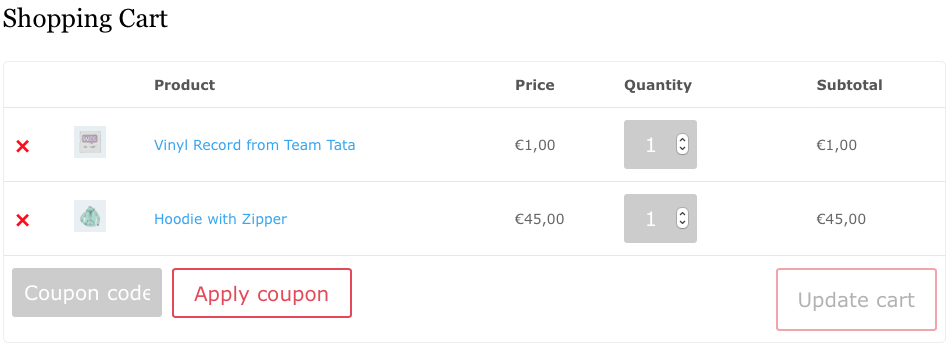}
    \caption{\emph{Sneak} into basket approach}
    \label{fig:sneaking}
\end{figure}

\noindent
\textbf{\emph{Sneaking:}}
In addition to a confirmshaming pattern, our \textsc{dark} shopping cart embedded a sneaking pattern; a hidden action which is accompanied by potentially undesired effects or unexpected costs. In other words, if the user would be aware of the hidden procedure or recognize it, he/she would object \cite{gray2018dark}. In our case, the \textsc{dark} version of our online shop automatically added a Vinyl Record for \EUR 1.00 to a user's shopping basked (cf. Fig.\ref{fig:sneaking}) whereas the \textsc{clean} version did not.

\begin{figure}[ht]
    \centering
    \includegraphics[width=\linewidth]{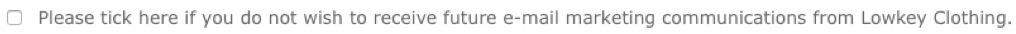}
    \caption{\emph{Double Negative} checkbox used for the newsletter subscription}
    \label{fig:DN}
\end{figure}

\noindent
\textbf{\emph{Double Negative:}} 
Concluding, our \textsc{dark} online shop also implemented a double negative pattern. During the check-out process users had to consent to terms and conditions as well as to the shops privacy policy. By doing this, we also asked for a potential newsletter subscription (cf. Fig.~\ref{fig:DN}). 
The used double negative formulation, as described by Gray et al.'s~\cite{gray2018dark} as aesthetic manipulation, asked users to tick the box in order to NOT receive the newsletter. The \textsc{clean} version of the online shop asked them to tick the box in case he/she wanted to receive future marketing communication.

\subsection{Questionnaires}\label{sec:questionnaire}
After completing their interaction with either the \textsc{clean} or the \textsc{dark} version of the online shop, participants were asked to fill in a questionnaire survey. Survey items focused on users' perceived annoyance level and respective brand trust. The level of perceived annoyance was measured overall (\textit{``Please rate your overall level of annoyance during your visit of the Lowkey Clothing webshop''}) and for each respective dark pattern (\textit{``Please rate your level of annoyance regarding pattern XX''}) on a 5-point Likert scale ranging from \textit{1=``not annoyed at all''} to \textit{5=``very annoyed''}.

As for brand trust, we used the Brand Trust Scale (BTS) by Delgado-Ballester and colleagues~\cite{delgado2003development}, which allows for quantitative measures based on two dimensions, i.e. \textit{fiability} and \textit{intentionality}. \textit{Fiability} focuses on need satisfaction and is thus related to a brand's (company's) value promise and whether a customer believes in it; i.e. whether the company sticks to its promises. \textit{Intentionality} focuses on hypothetical, unexpected problems with a product or brand and evaluates how a brand (company) deals with consumer interests. In other words, it represents the customer's emotional security towards a brand's (company's) problem-solving behaviour. Both dimensions consist of four questions, each measured on a 5-point Likert scale ranging from \textit{1=``completely disagree''} to \textit{5=``completely agree''} (cf. Table~\ref{tab:bts}).

\begin{table}[ht]
\caption{Question items on brand trust according to Delgado-Ballester et al.'s BTS scale~\cite{delgado2003development}.}
\label{tab:bts}
\centering
\begin{tabular}{p{15mm} p{105mm}}
\toprule
    \emph{Fiability} & \\
    \midrule
    BTS-F1 -- & {\itshape With Lowkey Clothing I did obtain what I looked for in an online shop.} \\
    BTS-F2 -- & {\itshape Lowkey Clothing was always at my consumption expectations level.} \\
    BTS-F3 -- & {\itshape Lowkey Clothing gave me confidence and certainty in the consumption of clothes.} \\
    BTS-F4 -- & {\itshape Lowkey Clothing never disappointed me so far.} \\
    \noalign{\vskip 2mm} 
    \midrule
    \emph{Intentionality} & \\
    \midrule
    BTS-I1 -- & {\itshape Lowkey Clothing seemed honest and sincere in its explanations.} \\
    BTS-I2 -- & {\itshape I could rely on Lowkey Clothing.} \\
    BTS-I3 -- & {\itshape Lowkey Clothing would make any effort to make me be satisfied.} \\
    BTS-I4 -- & {\itshape Lowkey Clothing would repay me in some way for a problem with the hoodie.} \\
    \noalign{\vskip 2mm} 
\bottomrule
\end{tabular}
\end{table}

In order to investigate potential experience effects, we used the \textit{Affinity Towards Technological Interaction} (ATI) scale~\cite{franke2019personal}. It acts as an indicator for effective interaction with technology and thus helps characterize user diversity. The ATI is grounded in the NFC (i.e., Need For Cognition) construct and consists of nine items, each measured on a 6-point Likert scale ranging from \textit{1=``completely disagree''} to \textit{6=``completely agree''} (cf. Table~\ref{tab:ati}). 

\begin{table}[ht]
\caption{Question items on technology affinity according to Franke et al.'s ATI scale~\cite{franke2019personal}.}
\label{tab:ati}
\centering
\begin{tabular}{p{15mm} p{105mm}}
\toprule
    \noalign{\vskip 2mm} 
    ATI1 -- & {\itshape I like to occupy myself in greater detail with technial systems.} \\
    ATI2 -- & {\itshape I like testing the functions of new technical systems.} \\
    ATI3 -- & {\itshape I predominantly deal with technical systems because I have to.} \\
    ATI4 -- & {\itshape When I have a new technical system in front of me, I try it out intensively.} \\
    ATI5 -- & {\itshape I enjoy spending time becoming acquainted with a new technical system.} \\
    ATI6 -- & {\itshape It is enough for me that a technical system works; I don't care how or why.} \\
    ATI7 -- & {\itshape I try to understand how a technical system exactly works.} \\
    ATI8 -- & {\itshape It is enough for me to know the basic functions of a technical system.} \\
    ATI9 -- & {\itshape I try to make full use of the capabilities of a technical system.} \\
    \noalign{\vskip 2mm} 
\bottomrule
\end{tabular}
\end{table}

Finally, we collected basic demographic data (i.e., \emph{Gender}, \emph{Age}, \emph{Country of Residence} and \emph{Level of Education}) in order to validate our sample (cf. Section~\ref{sec:sampling}). 

\subsection{Ethics, Sampling and Study Period}\label{sec:sampling}
The study followed common ethical considerations concerning research with human participation. Respective approval was obtained from the school's Ethics Commission in May 2020. Study sampling focused on representatives from Generation Y, i.e. so-called \emph{digital natives} who were born between 1980 and 2000 (note: we are aware that the PEW Research Institute defines different age brackets for Generation Y\footnote{Online: https://www.pewresearch.org/fact-tank/2019/01/17/where-millennials-end-and-generation-z-begins/ [accessed: February 12\textsuperscript{th} 2021]}, yet did not consider this slight variation relevant with respect to our target group). Most of them have been growing up with technology and are thus accustomed to shopping online~\cite{jones2007future}\cite{Prensky2001}. We reached out to potential participants via social media, direct messaging, as well as face-to-face contact. The two versions of our online shop (i.e., \textsc{dark} and \textsc{clean}) as well as the above described questionnaires were available from May 20\textsuperscript{th} to July 6\textsuperscript{th} 2020, which amounts to a total study period of 47 days.

\subsection{Study Procedure}
\begin{figure}[ht]
    \centering
    \includegraphics[width=100mm]{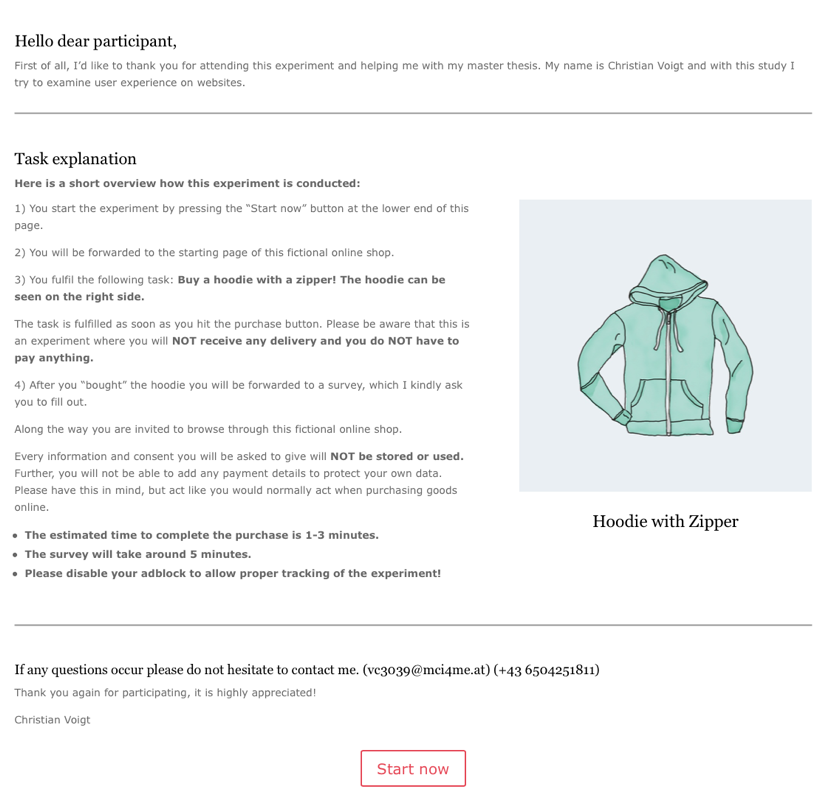}
    \caption{Introduction and task description}
    \label{fig:briefing_exp}
\end{figure}

Participants were first given some background information on the study goals (i.e., investigation of behaviour in online shopping) and then asked to consent to data processing (note: they were not told about the potential use of dark patterns). Next, they were given the task to buy a distinct product (i.e., a hoodie with a zipper) via the online shop version they were forwarded to (cf. Fig.~\ref{fig:briefing_exp}). 

\noindent
As mentioned earlier, approx. $2/3$ of respondents were forwarded to the \textsc{dark} version of the shop and $1/3$ to the \textsc{clean} one. Both versions were designed to appear as realistic as possible, without asking users to provide any payment details. After participants successfully completed the check-out process (sans payment), they were forwarded to the questionnaire survey. Its purpose was scientifically reasoned, yet its connection to our key survey components, i.e. \textit{perceived level of annoyance}, \textit{expressed brand trust} and \textit{affinity for technology}, was not further explained so as to inhibit bias. Still, our survey introduced participants to dark patterns and asked them to rate the level of annoyance they perceived with each of the patterns they were exposed to, before moving on to the questions on brand trust and affinity for technology. Note: with respect to the used \emph{Double Negative} and \emph{Sneaking} patterns, participants were first asked whether they noticed the pattern during their preceding shopping task. Only if they noticed the pattern, they were then asked to rate their perceived level of annoyance.  


\section{Results}\label{sec:results}
A total of $n=204$ participants (48.53\% female) completed the study, of whom $n_{dark}=134$ used the \textsc{dark} and $n_{clean}=70$ the \textsc{clean} version of our online shop. For the analyses we used a 95\% confidence level. 
Results show significant differences between the two versions of the online shop regarding both perceived annoyance and expressed brand trust (cf. Table~\ref{tab:evalbrandtrust}). On the one hand, participants who used the \textsc{dark} version reported a higher annoyance level than those who used the \textsc{clean} version. 
On the other hand, the expressed brand trust was lower with those participants who used the \textsc{dark} version than those who used the \textsc{clean} version. Such applied to both the \textit{fiability}  as well as the \textit{intentionality} dimension.

\begin{table}
\caption{Differences between the \textsc{dark} and the \textsc{clean} version of the online shop concerning perceived level of annoyance and expressed brand trust; cf. Section~\ref{sec:questionnaire} for further details on the used BTS scale.}
\label{tab:evalbrandtrust}
\centering
\begin{tabular}{ l  c  c  c }
\toprule
     & \textsc{dark}\enspace & \textsc{clean}\enspace & $p$\\
\midrule
    $Mean_{Annoyance}$\enspace & $3.44$\enspace & $2.34$\enspace & $0.000$ \\
    $SD_{Annoyance}$\enspace & $1.173$\enspace & $1.178$\enspace & \\
\midrule
    $Mean_{Fiability}$\enspace & $3.26$\enspace & $3.55$\enspace & $0.011$ \\
    $SD_{Fiability}$\enspace & $0.789$\enspace & $0.727$\enspace & \\
\midrule
    $Mean_{Intentionality}$\enspace & $3.07$\enspace & $3.42$\enspace & $0.003$ \\
    $SD_{Intentionality}$\enspace & $0.838$\enspace & $0.679$\enspace & \\
\bottomrule
\end{tabular}
\end{table}


\noindent
In addition, only 24\% of users from the \textsc{clean} version reported negative experiences, whereas more than 70\% of users from the \textsc{dark} version highlighted unintended or negative effects. When asking participants what triggered these negative experience, over 75\% recalled the returning advertisement pop-ups (i.e., \emph{Nagging}) as being particularly annoying. 
The data furthermore shows a connection between participants' overall level of annoyance and their brand trust. It points to a statistically significant correlation between the overall level of annoyance and the fiability dimension ($r_{s}=-.457; r^2=.2970; p=.000$), as well as the intentionality dimension ($r_{s}=-.545; r^2=.2088; p=.000$). 
A connection between participants' affinity for technology use and their ability to detect dark patterns, however, is not supported by the collected data (recognition of \emph{Double Negative}: $p=0.215$; recognition of \emph{Sneaking}: $p=0.232$), which may underline the often rather subconscious nature of these interface elements. 
Finally, regarding the annoyance level of perceived dark patterns our data shows that participants found \emph{Sneaking} to be most displeasing, followed by \emph{Confirmshaming} and \emph{Nagging} (cf. Table~\ref{tab:evaluation}). 



\begin{table}
\caption{Perceived level of annoyance regarding each of the experienced dark patterns; Scale: \textit{``1=not annoyed at all''} to \textit{15=``very annoyed''}}
\label{tab:evaluation}
\centering
\begin{tabular}{ l  c  c  c  c  c }
\toprule
     & Forced Action\enspace & Nagging\enspace & Confirmshaming\enspace & Sneaking\enspace & Double Negative \\
\midrule
    $Mean$ & 3.16 & 4.02 & 4.28 & 4.53 & 3.53  \\
    $SD$ & 1.268 & 1.007 & 1.127 & 0.822 & 0.492  \\
\bottomrule
\end{tabular}
\end{table}

\section{Discussion and Limitations}\label{sec:discussion}
Our study results indicate that dark patterns do harm brand trust and that they are connected to an increased level of annoyance perceived by customers when shopping online. Yet, the results also show that dark patterns do work. The aesthetic manipulation via \emph{Double Negative} used by the \textsc{dark} version of our online shop increased newsletter subscriptions by 12\% compared to the \textsc{clean} version. Surprisingly, those 60\% of participants who did recognize the manipulation, rated its level of annoyance rather low (i.e., $Mean=3.53$). Also the \emph{Sneaking} pattern worked in 38 out of 134 cases. This means, that over 11\% of our study participant accidentally purchased an additional product; even though this dark pattern was rated as highly annoying by over 68\% of all the participants who were exposed to the \textsc{dark} version of the shop. 

Considering our results, we need to highlight that respective ratings only apply to those five distinct patterns applied in our study. They may not generalise to other instances, that potentially use very different forms and appearances. For example, the annoyance of \emph{Forced Action} was rated rather low in our study. This may be caused by the implementation via a cookie banner when first entering the website. Today, users face cookie banners on nearly every website and may thus be more forgiving. On the other hand, \emph{Sneaking} is a less commonly used pattern, for which it may have triggered higher annoyance ratings in our study. 

As for the measured \textit{brand trust}, our results may also be considered limited, as our setup used a fictional brand: \emph{Lowkey Clothing}. The BTS scale aims to measure trust in brands that are already known to a consumer~\cite{delgado2003development}. In our study however, participants rated trust based solely on the experience gained from this one-off shopping task, which may certainly be too little time to build up a sufficient trust level.

Finally, our data does not point to a connection between participants' ATI scores and their ability to recognize dark patterns. This is unexpected, since one would assume that through a higher affinity towards technological interaction, one would be more aware of malicious design and therefore recognize dark patterns more often. Furthermore, the data does not support a connection between participants' ATI scores and their perceived level of annoyance.
One would assume at least a weak link, since users with a higher affinity towards technological interaction tend to spend more time online, and may thus be confronted more often with dark pattern. The empirical lack thereof underlines the importance of our study and shows that more research is necessary in order to increase our understanding of the effects dark patterns have on users.
\section{Conclusion and Future Outlook}\label{sec:conclusion}
We conducted an experiment-based survey to understand a potential connection between the use of dark patterns in online shopping, users’ level of perceived annoyances, and their expressed brand trust. While results support connections to both (i.e., perceived annoyance and expressed brand trust), its impact is not yet fully understood. Hence, future work should aim to model these connections so that we may be able to identify a threshold at which the negative effects with respect to annoyance and brand trust outweigh the gains resulting from the use of dark patterns. In addition, we see a need for targeted information and user education. Although the results of our study rejected the assumption that affinity for technology yields a higher level of dark pattern recognition, we do believe that more targeted awareness raising programs may still help users recognize hidden tricks and consequently prevent unintended actions.   

To our knowledge, there has been no empirical analysis investigating the connection between dark patterns and brand trust so far. And although our results are still preliminary and require further validation, they underline the negative side effects of dark patterns. Hence, companies may reflect on the use of dark pattern or at least counterbalance its value against the detrimental effects it could bring to ones' perceived brand value. 



\bibliography{hcii}
\bibliographystyle{splncs04}
\end{document}